\documentclass[aps,prl,superscriptaddress,nofootinbib,floatfix]{revtex4}

\pdfoutput=1

\usepackage{epsfig}
\usepackage{bm}
\usepackage{amssymb}
\usepackage{amsmath}
\usepackage{color}
\usepackage{subfigure}

\newcommand{\be}{\begin{equation}}
\newcommand{\ee}{\end{equation}}
\newcommand{\bea}{\begin{eqnarray}}
\newcommand{\eea}{\end{eqnarray}}
\newcommand{\bfk}{\mathbf{k}}

\newcommand{\bfp}{\mathbf{p}}
\newcommand{\knn}{\textrm{Kn}}

\usepackage{amsmath}
\usepackage{amssymb}
\usepackage{amsthm}
\usepackage{mathrsfs}
\usepackage{graphicx}
\usepackage{epstopdf}
\usepackage{fancyhdr}
\usepackage{array}
\usepackage[all]{xy}
\usepackage{eufrak}
\usepackage{euscript}
\usepackage{enumerate}
\usepackage{slashed}
\usepackage{hyperref}
\usepackage{subfigure}
\usepackage{epstopdf}

\begin{document}

\title{Exact hydrodynamic attractor of an ultrarelativistic gas of hard spheres}
\date{\today}

\author{Gabriel S.~Denicol}
\affiliation{Instituto de F\'isica, Universidade Federal Fluminense, UFF, Niter\'oi, 24210-346, RJ, Brazil}
\email{gsdenicol@id.uff.br}

\author{Jorge Noronha}
\affiliation{Department of Physics, University of Illinois, 1110 W. Green St., Urbana IL 61801-3080, USA}
\email{jn0508@illinois.edu}

\begin{abstract}
We derive the general analytical solution of the viscous hydrodynamic equations for an ultrarelativistic gas of hard spheres undergoing Bjorken expansion, taking into account effects from particle number conservation, and use it to analytically determine its attractor at late times. Differently than all the cases considered before involving rapidly expanding fluids, in this example the gradient expansion converges. We exactly determine the hydrodynamic attractor of this system when its microscopic dynamics is modeled by the Boltzmann equation with a fully nonlinear collision kernel. The exact late time attractor of this system can be reasonably described by hydrodynamics even when the gradients are large.       
\end{abstract}

\maketitle


\noindent \textsl{1. Introduction.} Hydrodynamics is an effective theory describing the dynamics of a many-body system at times and distances that are considerably larger than any microscopic scale. Such a separation of scales is commonly characterized by a dimensionless quantity called the Knudsen number, $\textrm{Kn}=\ell/L$, where $\ell$ is the relevant microscopic scale, and $L$ is a macroscopic scale associated with the gradients of conserved currents. Since the seminal work of Hilbert, Chapman, and Enskog \cite{chapmanenskog,cercignani}, hydrodynamic descriptions have been constructed via truncations of a systematic expansion of the conserved currents in powers of the Knudsen number. The zeroth-order truncation of this gradient expansion leads to the equations of ideal fluid dynamics, while Navier-Stokes theory corresponds to its first-order truncation \cite{LandauLifshitzFluids}.

The large order behavior of the gradient expansion is generally not well understood and, so far, has only been studied in a few problems that involve highly symmetrical, relativistically expanding systems \cite{Heller:2013fn}. In such examples, the series was shown to diverge \cite{Heller:2013fn,Heller:2015dha,Buchel:2016cbj,Denicol:2016bjh,Heller:2016rtz} and, hence, in these cases the gradient expansion cannot be used to systematically define and improve hydrodynamic formulations. This motivated the search for alternative ways to define hydrodynamics in such a way that it could also be consistent (and accurate) even when the Knudsen number is not parametrically small \cite{Romatschke:2017vte,Spalinski:2017mel,Strickland:2017kux,Romatschke:2017acs,Florkowski:2017olj,Bemfica:2017wps,Denicol:2017lxn,Behtash:2017wqg,Blaizot:2017ucy,Casalderrey-Solana:2017zyh,Heller:2018qvh,Almaalol:2018ynx,Denicol:2018pak,Casalderrey-Solana:2018uag,Behtash:2018moe,Behtash:2019txb,Strickland:2018ayk,Strickland:2019hff,Jaiswal:2019cju,Kurkela:2019set}.

The current working hypothesis is that hydrodynamics may be defined as a universal attractor \cite{Heller:2015dha} where the dissipative currents display universal behavior that is independent of their initial conditions. This indicates the emergence of constitutive relations involving the dissipative currents and the gradients of conserved quantities, which may be non-perturbative in the Knudsen number \cite{Heller:2015dha,Denicol:2016bjh,Heller:2016rtz,Denicol:2017lxn} and, therefore, cannot always be expressed in the form of a simple power series. Attractor solutions are rare and have only been obtained in simple systems (i.e., conformal limit and/or simplified kinetic models) under highly symmetrical expansion dynamics (i.e., Bjorken \cite{Bjorken:1982qr} and Gubser flows \cite{Gubser:2010ze}). 

This new understanding of the onset of hydrodynamics is particularly important in the theoretical description of the rapidly expanding hot and dense matter produced in ultrarelativistic heavy-ion collisions, where a fluid dynamical approximation appears to be valid even though the Knudsen number is large \cite{Niemi:2014wta,Noronha-Hostler:2015coa}. Interest in this open problem has been renewed since the observation of hydrodynamic signatures even in the extremely small and explosive systems produced in (moderately) high multiplicity proton-proton, proton-nucleus, and deuteron/helium-nucleus collisions \cite{Bozek:2011if,Weller:2017tsr,PHENIX:2018lia}. If hydrodynamic behavior can indeed emerge even far from equilibrium (where the Knudsen number is large) in the form of a universal attractor, this would naturally explain the puzzling observations found in these hadronic collisions.

In this paper we investigate the emergence of attractor behavior in an ultrarelativistic gas of hard spheres undergoing Bjorken expansion. In this system, we prove that the Knudsen number is constant, which makes it possible to extract the attractor solution in different regimes. We first show that the attractor solution can be determined analytically in Israel-Stewart-like theories derived from the Boltzmann equation. In this case, we further demonstrate that the gradient expansion converges absolutely in a finite range of Knudsen numbers -- this is the first example where such a series converges in an expanding system,  which changes the current view that the gradient expansion is always an asymptotic series. We then determine for the first time the hydrodynamic attractor of this system when its microscopic dynamics is described by the integro-differential Boltzmann equation with nonlinear collision kernel (assuming classical statistics), going beyond previous attractor analyses of kinetic models that were restricted to the relaxation time approximation. We show that both attractors agree when the Knudsen number is small, but even as the Knudsen number becomes large the deviations are never parametrically large.


\noindent \textsl{2.  General properties.} We consider a \textit{homogeneous} ultrarelativistic gas of hard spheres in Milne coordinates $x^\mu=(\tau,x,y,\varsigma)$ with line element (we set $\hbar=c=k_B=1$)
\begin{equation}
ds^{2}=g_{\mu \nu }dx^{\mu }dx^{\nu }=d\tau ^{2}-\left( dx^{2}+dy^{2}+\tau
^{2}d\varsigma ^{2}\right) \,,  \label{metric}
\end{equation}%
where $\tau =\sqrt{t^{2}-z^{2}}$ and $\varsigma =\tanh ^{-1}(z/t)$. We further assume that the system is
invariant under reflections around the longitudinal $\varsigma $-axis, which
restricts any normalized time-like 4-vector to be of the form $u^{\mu
}=\left( 1,0,0,0\right) $, while space-like 4-vectors must vanish. In Cartesian coordinates this corresponds to a system undergoing longitudinal expansion called Bjorken flow \cite{Bjorken:1982qr}. 

Under these assumptions the energy-momentum tensor $T^{\mu\nu}$ is diagonal and can be written as 
\be
T^{\mu}_\nu = \textrm{diag}\left(\varepsilon,P-\pi/2,P-\pi/2,P+\pi\right)
\label{defineTmunu}
 \ee 
where $\varepsilon(\tau)$ is the energy density, $P=\varepsilon/3$ is the thermodynamic pressure of the ultrarelativistic gas, and $\pi(\tau)$ is the longitudinal component of the shear stress tensor. We remark that in this system the bulk viscous pressure vanishes. We shall only take into account binary collisions, which implies that particle number is conserved. Under Bjorken flow, the associated particle 4-current is simply given by $N^\mu = (n,0,0,0)$, where $n=P/T$ is the particle density and $T$ is the temperature.

A key feature of Bjorken flow with particle number conservation is that the dynamics of the particle density decouples from that of the energy-momentum tensor. The conservation law, $\nabla_\mu N^\mu=0$, leads to a simple equation,
\be
\frac{dn}{d\tau} + \frac{n}{\tau}=0,
\ee
whose solution can be determined analytically regardless of the microscopic properties of the fluid. One finds
\be
n(\tau) = n_0 \tau_0/\tau,
\ee
where $n_0$ is the particle density at a time $\tau_0$. This feature allows us to determine analytically the mean free path of the gas of hard spheres, $\ell_\textrm{mfp} = 1/(n\sigma_T)$, which grows linearly with $\tau$. 

The mean free path determines the microscopic scale of the gas and its linear dependence with $\tau$ dramatically affects the properties of the gradient expansion. This happens because the macroscopic scale associated with the gradients of hydrodynamic variables also goes as $\tau$. Therefore, the Knudsen number, which is given by 
\be
\textrm{Kn}=\frac{\ell_\textrm{mfp}}{\tau} = \frac{1}{n_0 \tau_0\sigma_T},
\label{defineKn}
\ee
is a constant in Bjorken flow. Hence, we can treat the Knudsen number as a tunable parameter and study new situations where the Knudsen number is large but there are no transient effects. 

This behavior is very different than the one found in conformal systems \cite{Baier:2007ix,Marrochio:2013wla,Denicol:2014xca,Denicol:2014tha}, which have been thoroughly studied before \cite{Heller:2015dha}. In this case, the microscopic scale goes as $\sim 1/T$ and the Knudsen number is time dependent since $\textrm{Kn}  \sim 1/(T\tau)$, being thus very large at early times while it decreases as $\sim \tau^{-2/3}$ when $\tau$ is large\footnote{We remark in passing that the dimensionless time variable $w=T\tau$ used in previous studies \cite{Heller:2015dha} is proportional to the inverse Knudsen number in Bjorken flow.}. This means that a conformal system undergoing Bjorken expansion evolves towards local equilibrium, and the hydrodynamic attractor of such systems is always well described by relativistic Navier-Stokes theory at sufficiently late times. Furthermore, it also means that large Knudsen numbers can only be probed at early times where transient effects are expected to be dominant. In this sense,  going beyond this idealized situation is important to understand the cases where the asymptotic regime remains far from equilibrium, as it occurs in the dynamics of the quark-gluon plasma formed in ultrarelativistic heavy-ion collisions. The simplest microscopic system where this can be systematically studied is the one considered in this paper. 

\noindent \textsl{3.  Analytical solution of viscous hydrodynamics.} We start our analysis by investigating the hydrodynamic regime of Israel-Stewart-like theories \cite{MIS-6,Muronga:2003ta,Denicol:2012cn} derived from the Boltzmann equation. 
In Bjorken flow, energy and momentum conservation, $\nabla_\mu T^{\mu\nu}=0$, is encoded in a single equation for the energy density, 
\be
\frac{1}{\varepsilon}\frac{d\varepsilon}{d\tau}+\frac{4}{3\tau}=\frac{4\chi}{3\tau},
\label{tempeq}
\ee
where we introduced the dimensionless quantity $\chi = \pi/(4P)$. This conservation law is supplemented by a dynamical equation for $\chi$ derived from kinetic theory \cite{Denicol:2012cn,Denicol:2012es,Denicol:2014vaa}, which reads
\be
\tau_\pi\frac{d\chi}{d\tau} + \frac{4\tau_\pi}{3\tau}\chi^2 + \frac{\tau_{\pi\pi}}{3\tau}\chi+ \chi = \frac{\eta}{3P\tau}.
\label{eqchi}
\ee
Above, the shear viscosity $\eta $, relaxation time $\tau_\pi$, and $\tau_{\pi\pi}$ are computed from the Boltzmann equation \cite{Denicol:2012cn,Denicol:2012es,Denicol:2014vaa} and can be expressed in the following simple form 
\be
\eta = a\frac{T}{\sigma_T}, \qquad \tau_\pi=b\,\frac{\eta}{4P}, \qquad \tau_{\pi\pi}=3\lambda\, \tau_\pi,
\label{definetransport}
\ee
where $\sigma_T$ is the total cross section and $a$, $b$, and $\lambda$ are constants. In calculations in kinetic theory within the 14-moment approximation, one finds $a=4/3$, $b=5$, and $\lambda=10/21$. Nevertheless, the results derived in this section are valid for any positive values of these coefficients.

Using the transport coefficients from \eqref{definetransport}, the equation for $\chi$ becomes 
\be
\knn\left( \tau\frac{d\chi}{d\tau} + \frac{4}{3} \chi^2+ \lambda  \chi\right) + \frac{4}{ab}\chi= \frac{4}{3b}\knn.
\label{eqchinew}
\ee
One can see that $\tau d\chi/d\tau$ depends solely on $\chi$ since $\textrm{Kn}$ is constant, which allows us to analytically solve this equation. The analytical solution of \eqref{eqchinew} is
\be
\chi(\tau) = \mathcal{A}\left[\frac{1-c_0\left(\tau_0/\tau\right)^{8\mathcal{A}/3}}{1+c_0\left(\tau_0/\tau\right)^{8\mathcal{A}/3}}\right]-\frac{3}{8}\frac{\left(4+ a b\lambda\,\knn\right)}{ab\,\knn},
   \label{foda1}
   \ee
where 
\be
\mathcal{A} = \frac{\sqrt{144 + ab\,\knn \left(72\lambda+a\,\knn\left(64+9 b \lambda^2\right)\right)}}{8 ab \,\knn}
\ee
and $c_0$ is an integration constant. The energy density can then be determined to be  
   \bea
\varepsilon(\tau) &=& c_1  \left(\frac{\tau_0 }{\tau}\right)^{\frac{4}{3}\left(1-\mathcal{A}\right)+\frac{\lambda}{2}+\frac{2}{ab\,\knn}} \left[1+c_0\left(\frac{\tau_0}{\tau}\right)^{8\mathcal{A}/3}\right],\nonumber
\eea
where $c_1 = \varepsilon(\tau_0)/(1+c_0)$ is another integration constant.

At sufficiently late times, we see that \eqref{foda1} becomes independent of its initial condition and $\chi$ approaches the \emph{constant} value 
\be
\chi_{\textrm{att}} =\mathcal{A} -\frac{3}{8}\frac{\left(4+ a b\lambda\,\knn\right)}{ab\,\knn}.
\label{atratoranal}
\ee
We remark that this asymptotic value is also a solution of \eqref{eqchinew}. Therefore, \eqref{atratoranal} is actually an \emph{attractor solution} where the energy density is exactly given by
\be
\varepsilon_{\textrm{att}}(\tau) =c_1  \left(\frac{\tau_0 }{\tau}\right)^{\frac{4}{3} (1-\chi_{\textrm{att}})}.
\ee
Note that all the information about the initial values are carried by the energy density. While $\chi_{\textrm{att}}$ is constant in time, we remark that it is a nontrivial function of the Knudsen number and, in principle, all the properties of the gradient expansion can  be extracted from \eqref{atratoranal}. Furthermore, we stress that here the 0th-order approximation of the slow-roll series \cite{Heller:2015dha,Denicol:2017lxn,Denicol:2018pak} gives the exact result for the attractor. 

In the gradient expansion, the attractor is expressed as a power series in Knudsen number. While the expression for the analytical attractor in \eqref{atratoranal} is formally singular at $\knn=0$, we note that this is a removable singularity\footnote{For example, the function $\frac{\sin x}{x}$ is also formally singular at $x=0$. However, this singularity is removable and an analytic continuation, which is regular at $x=0$, can be defined for this function using the series $\sum_{n=0}^\infty (-1)^n x^{2n}/(2n+1)!$.} since $\lim_{\knn\to 0} \chi_{\textrm{att}}=0$. As a matter of fact, one can expand the square root in \eqref{atratoranal} in powers of $\knn$ to define a series that provides an analytical continuation of the attractor that is regular at the origin. For the sake of simplicity, we focus on the case where $\lambda=0$, which leads to the series
\be
\chi_{\textrm{att}} =  \,\sum_{n=1}^\infty \chi_{n}\,\textrm{Kn}^{n},
\ee
where the expansion coefficients are
\be
\chi_{n} =a \left(\frac{2}{3}\right)^n \binom{\frac{1}{2}}{\frac{n+1}{2}}
   \left(a^2 b\right)^{\frac{n-1}{2}}.
\ee
This series representation converges absolutely when $|\knn|<3/(2a\sqrt{b})$. This provides the first example where the gradient expansion can be shown to converge in a relativistically expanding system. As expected, the first nonzero term of the series is given by the relativistic Navier-Stokes value $a\textrm{Kn}/3$ \cite{LandauLifshitzFluids}. Furthermore, we note that if one takes the values of $a$ and $b$ calculated in kinetic theory in the 14-moment approximation, i.e., $a=4/3$ and $b=5$, the series converges absolutely when $|\textrm{Kn}| <9/(8\sqrt{5})\approx 0.5$. We have checked that similar conclusions are valid when $\lambda \neq 0$ and that the series in this case converges absolutely when $|\knn| < 12/[a(\sqrt{18 b^2 \lambda ^2+64 b}+3  b \lambda)]$.

It was generally expected that the gradient expansion diverges, at least in the case of rapidly expanding systems. However, an important exception to a general statement regarding the divergent nature of the gradient series has recently come to light from in-depth studies of holographic fluids in  \cite{Withers:2018srf} and \cite{Grozdanov:2019kge,Grozdanov:2019uhi}, which have shown that the near-equilibrium, linearized spatial gradient expansion converges. A similar problem was also investigated a long time ago in the context of kinetic theory in \cite{1965PhFl....8.1580M}. Based on these results, the authors of Refs.\ \cite{Grozdanov:2019kge,Grozdanov:2019uhi} argued that a vanishing radius of convergence should not be interpreted as a general property of the gradient expansion. Rather, they suggested that its occurrence in previous examples may only reflect universal \emph{singular} features of the flow displayed by rapidly expanding systems. In this paper we show that this may not be the case since the attractor derived above can be expressed as a convergent series, even though the system is rapidly expanding. Therefore, the fact that the system is expanding is not a sufficient\footnote{Nor is expansion a necessary condition for the divergence of the series. The gradient expansion was already shown to diverge even in stationary regimes, see Ref.\ \cite{PhysRevLett.56.1571}.} condition to guarantee that the gradient series diverges.

We emphasize that in this system the reliability of the Navier-Stokes limit is determined by the choice of the initial particle density $n_0$, initial time $\tau_0$, and the value of the total cross section $\sigma_T$, which determine $\textrm{Kn}$. If the Knudsen number is sufficiently small, the late time dynamics of $\chi$ will be approximated by Navier-Stokes constitutive relations. If $\textrm{Kn}$ is not small, but is still within the radius of convergence calculated above, the attractor can be systematically approximated by higher order truncations of the gradient series. On the other hand, if $\textrm{Kn}$ is outside the radius of convergence, the attractor given by \eqref{atratoranal} cannot be approximated by the usual constitutive relations associated with hydrodynamic behavior. The latter is a clear example of the emergence of hydrodynamics, i.e., the existence of universal constitutive relations between dissipative currents and gradients, even when the system is far from equilibrium.  

\begin{figure}
\includegraphics[width=0.6\textwidth]{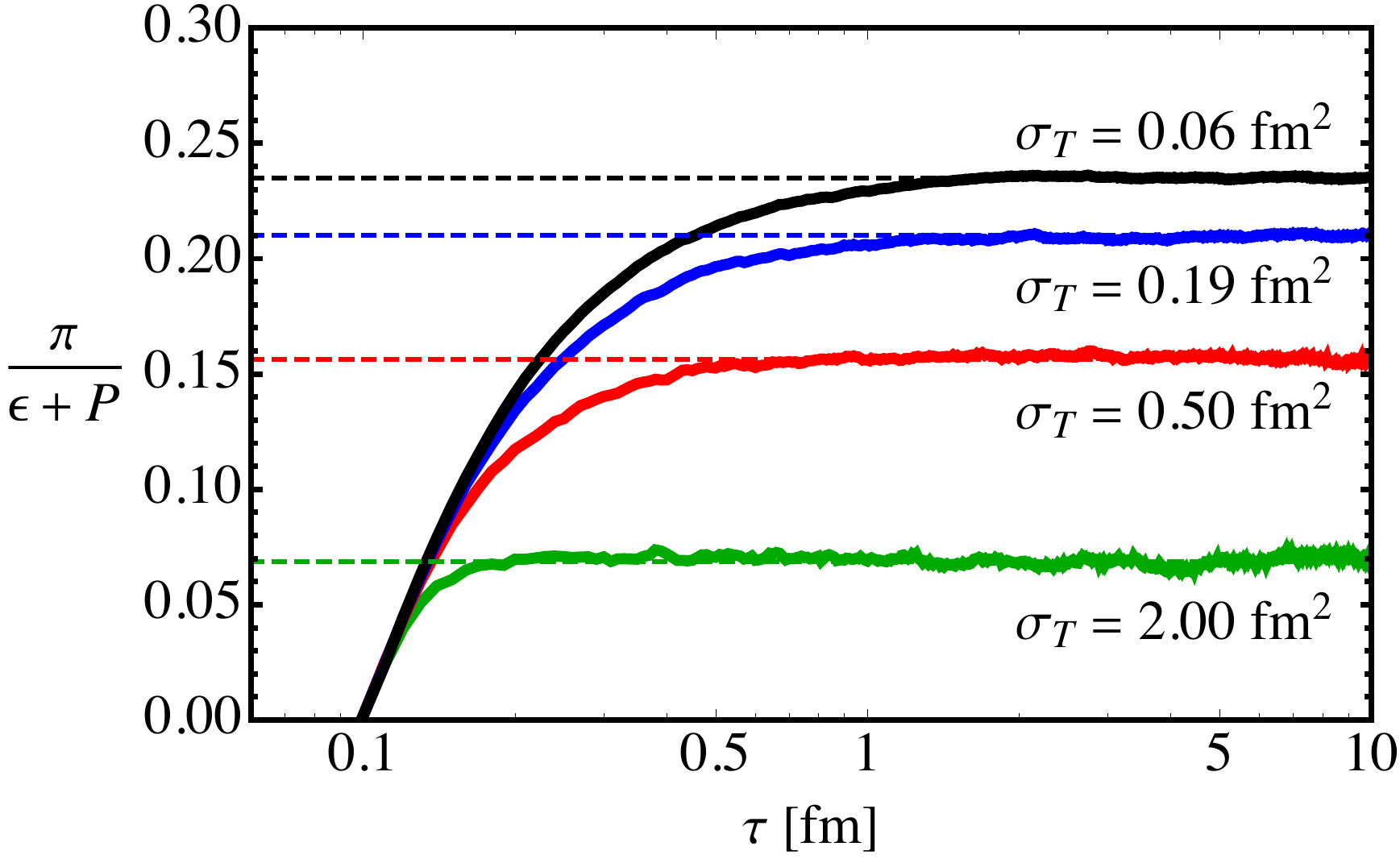}
\caption{Dynamical evolution of $\pi/(\varepsilon+P)$, computed using the Boltzmann equation \eqref{Boltzmanneq}, considering different values of the total cross section. The dashed curves indicate the asymptotic values of this quantity, which determine the attractor.}
\label{fig1}
\end{figure}


\noindent \textsl{4. Hydrodynamic attractor of the Boltzmann equation.} In this section we obtain the attractor of an ultrarelativistic gas of hard spheres now from a microscopic perspective using the Boltzmann equation \cite{degroot}. For Bjorken flow described in Milne coordinates \eqref{metric}, the contribution from the Christoffel symbols cancel exactly, and the Boltzmann equation becomes \cite{Denicol:2014xca,Denicol:2014tha}
\be
k_0 \partial_\tau f_\bfk  = \mathcal{C}[f,f],
\label{Boltzmanneq}
\ee
where $f_\bfk=f(\tau;k_0,k_\varsigma)$ is the single particle distribution function, $k_{\mu }=(k_{0},\mathbf{k})$ is the covariant 4-momentum of the massless particles with $k_0=\sqrt{k_{x}^{2}+k_{y}^{2}+k_{\varsigma
}^{2}/\tau ^{2}}$. The nonlinear collision kernel (assuming classical statistics) reads \cite{Bazow:2015dha,Bazow:2016oky}
\be
\mathcal{C}[f,f] = \frac{1}{2}\int_{k',p,p'}W_{\bfk\bfk'\to \bfp\bfp'}\left(f_\bfp f_{\bfp'}-f_\bfk f_{\bfk'}\right),
\ee
where $\int_k = \int d^3\bfk/\left[(2\pi)^3 \tau k_0\right]$, and $W_{\bfk\bfk'\to \bfp\bfp'}$ is the transition rate given by
\be
W_{\bfk\bfk'\to \bfp\bfp'} = \tau\, \sigma_T \,s\, (2\pi)^5\delta^{(4)}(k_\mu+k'_\mu-p_\mu-p'_\mu),
\ee
with Mandelstan variable $s = 2k^\mu k'_\mu$. Given a solution of the Boltzmann equation, one can determine any of the moments of the distribution function, such as the particle number density $n= \int_k k_0 f_\bfk$, the energy density $\varepsilon = \int_k k_0^2 f_\bfk$, and the shear stress tensor determined by $\pi=\int_k \left[k_0^2/3-\left(k_{\varsigma }/\tau \right) ^{2 }\right]f_{\mathbf{k}}$. 

\begin{figure}
\includegraphics[width=0.6\textwidth]{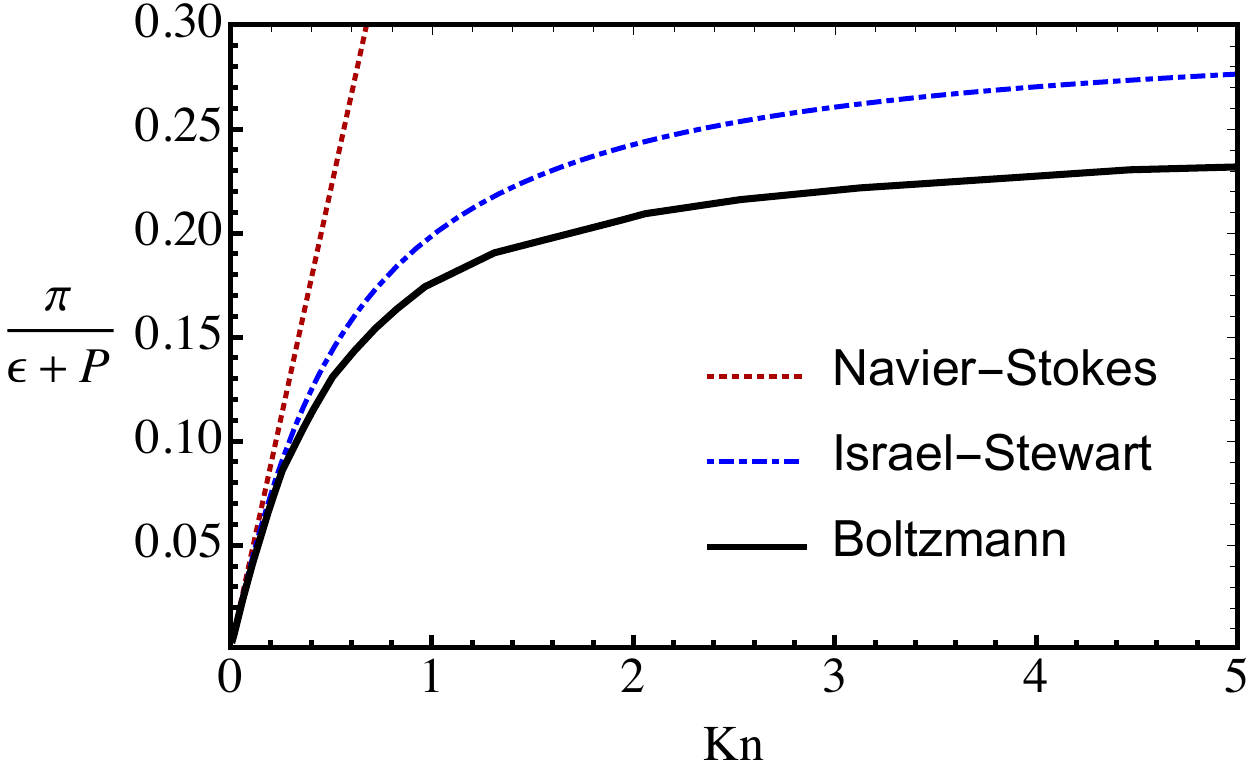}
\caption{Attractor solutions for $\pi/(\varepsilon+P)$ as a function of the Knudsen number for the Boltzmann equation (solid black curve) and for Israel-Stewart theory computed using the 14-moment approximation (dot-dashed blue curve). For a comparison, we also show the Navier-Stokes constitutive relation (red dotted curve).}
\label{fig2}
\end{figure}

We note that the spatially homogeneous Boltzmann equation in Milne coordinates in \eqref{Boltzmanneq} can be mapped onto a homogeneous Boltzmann equation in Cartesian coordinates defined in a volume that increases linearly with time, and with a longitudinal momentum that decreases with time as $k_\varsigma/\tau$. Using this, we solved \eqref{Boltzmanneq} numerically employing the method outlined in Ref.\ \cite{Xu:2004mz}. We checked that our results for a conformal system (with $\sigma_T \sim 1/T^2$) reproduce the numerical solutions of the Boltzmann equation in Bjorken flow previously performed in Ref.\ \cite{El:2009vj}.  

Since the Knudsen number is constant, the attractor solution for $\chi$ obtained from the Boltzmann equation must also be constant. As a matter of fact, in this system any dimensionless quantity constructed using moments of the Boltzmann distribution must asymptote to a constant. Therefore, the attractor for $\chi$ can be directly extracted from the numerical solution of the Boltzmann equation evaluated at sufficiently late times. One can see this in practice in Fig.\ \ref{fig1}, where the dynamical evolution of $\chi$, computed using the Boltzmann equation, is shown for different values of total cross section. We show several simulations that start in equilibrium at $\tau_0 = 0.1$ fm and $T(\tau_0)=0.5$ GeV (at vanishing chemical potential), with $\sigma_T \in [0.06,2]$ fm$^{2}$. Therefore, each simulation has a different value of Knudsen number. It is clear that $\chi$ approaches a constant value at late times, which decreases with increasing $\sigma_T$ and, hence, decreasing $\textrm{Kn}$. As a check, we also considered a case where $\tau_0=1$ fm, with the same initial temperature and chemical potential as before, and verified that the constant value of $\chi$ achieved in this case is the same as the value found  when the cross sections are ten times smaller. This shows that the attractor solution for the Boltzmann equation solely depends on $\textrm{Kn}$.

In Fig.\ \ref{fig2}, we plot the exact Boltzmann attractor solution for $\chi$ (solid black curve) as a function of $\textrm{Kn}$. This result is compared to the analytical attractor solution in Eq.\ \eqref{atratoranal} (dot-dashed blue curve) calculated with transport coefficients given by the 14-moment approximation. For the sake of illustration, we also plot the Navier-Stokes approximation for the attractor (dotted red curve). Both attractor solutions display similar qualitative features: a linear behavior at small $\knn$, well described by Navier-Stokes theory, which eventually saturates to a given value when $\knn$ is large. Quantitatively, the attractor solutions agree when the Knudsen number is small, $\knn \leq 0.1$, but even as the Knudsen number becomes large the deviations are never parametrically large. As a matter of fact, we note that even when $\knn \gg 1$ the attractors never differ by more than 20\%. Such a small difference between the attractor solution found within the effective hydrodynamic description (Israel-Stewart theory), and the corresponding attractor of the full nonlinear Boltzmann equation, is remarkable. This result suggests that the effective hydrodynamic theories used in heavy-ion collision simulations \cite{Romatschke:2017ejr} can be extremely robust even when the Knudsen number is large.

It would be interesting to check if the Boltzmann attractor admits a convergent power series representation, as happened in the case of Israel-Stewart theory. However, implementing the gradient expansion for the Boltzmann equation, with the full collision operator, is a very complex task that has never been performed even in highly symmetric setups. So far, this has only been worked out using a toy model of the collision operator, known as the relaxation time approximation (RTA) \cite{degroot}, in Refs.\ \cite{Denicol:2016bjh,Heller:2016rtz}. Using the method described in \cite{Denicol:2016bjh}, we can at least demonstrate that the gradient expansion diverges for an ultrarelativistic gas of hard spheres described by the RTA Boltzmann equation. While this may indicate that the gradient series also diverges in the case of the full Boltzmann equation, this is certainly not a proof and, thus, remains as an open question in the field.


\noindent \textsl{4. Conclusions.} In this work we provided a comprehensive analysis of the far-from-equilibrium dynamics of an ultrarelativistic gas of hard spheres, described by the full Boltzmann equation, undergoing Bjorken flow. We demonstrated that the Israel-Stewart hydrodynamic equations, derived from the Boltzmann equation, can be solved analytically in this case. We then proved that an attractor solution does exist, and we determined it analytically. Differently than all the cases considered before involving rapidly expanding fluids, in this case the gradient expansion converges absolutely in a finite range of Knudsen numbers. This conclusively shows that the divergence of the gradient expansion is not a general property of rapidly expanding systems nor a singular feature of Bjorken flow. We further found, for the first time, an exact attractor of the full relativistic Boltzmann equation. We showed that the Boltzmann attractor and its Israel-Stewart counterpart qualitatively agree, with visible quantitative deviations appearing only at large Knudsen numbers, which however remain 20\% at best. This shows that hydrodynamic theories used in heavy-ion collision simulations can be very effective, even in the far-from-equilibrium regime where the Knudsen number is large. It is our hope that the results presented here may contribute towards understanding the surprising effectiveness of relativistic hydrodynamics in the description of the rapidly expanding quark-gluon plasma formed in hadronic collisions.

\section*{Acknowledgements} We thank U.\ Heinz who, after this work was completed and posted, made us aware of a talk by A.~Jaiswal at Hirschegg 2019 where the case of a constant Knudsen number was also reported. The authors thank Instituto de F\'isica da Universidade Estadual de Campinas (IFGW-UNICAMP) for its hospitality. GSD and JN thank Conselho Nacional de Desenvolvimento Cient\'ifico e Tecnol\'ogico (CNPq) for support. GSD thanks Funda\c c\~ao Carlos Chagas Filho de Amparo \`a Pesquisa do Estado do Rio de Janeiro (FAPERJ), grant number E-26/202.747/2018. JN acknowledges partial support from Funda\c c\~ao de Amparo \`a Pesquisa do Estado de S\~ao Paulo (FAPESP), grant number 2017/05685-2.

\bibliography{References_attractor.bib}

\end{document}